\documentclass[preprintnumbers,article,amsmath,amssymb,floatfix,10pt,prd,twocolumn,
superscriptaddress,nofootinbib]{revtex4-2}
\usepackage{bm}
\usepackage{amsfonts}
\usepackage{latexsym}
\usepackage[latin1]{inputenc}
\usepackage{graphicx}
\usepackage{amsmath}
\usepackage{palatino}
\usepackage{mathpazo}
\usepackage{textcomp}
\usepackage{float}
\usepackage{booktabs}
\usepackage{dcolumn}
\usepackage{ragged2e}
\usepackage{hyperref}
\hypersetup{colorlinks,citecolor=blue}
\hypersetup{colorlinks=true,linkcolor=red,filecolor=magenta,    urlcolor=blue}
\usepackage{xcolor}
\usepackage{orcidlink}
\usepackage{epsfig}
\usepackage{subfigure}
\usepackage{commath}

\def\jnl@style{\it}
\def\aaref@jnl#1{{\jnl@style#1}}

\def\aaref@jnl#1{{\jnl@style#1}}

\def\aj{\aaref@jnl{AJ}}                   
\def\apj{\aaref@jnl{ApJ}}                 
\def\apjl{\aaref@jnl{ApJ}}                
\def\apjs{\aaref@jnl{ApJS}}               
\def\apss{\aaref@jnl{Ap\&SS}}             
\def\aap{\aaref@jnl{A\&A}}                
\def\aapr{\aaref@jnl{A\&A~Rev.}}          
\def\aaps{\aaref@jnl{A\&AS}}              
\def\mnras{\aaref@jnl{Mon.~Not.~Roy.~Astron.~Soc.}}             
\def\prd{\aaref@jnl{Phys.~Rev.~D}}        
\def\prc{\aaref@jnl{Phys.~Rev.~C}}  
\def\prl{\aaref@jnl{Phys.~Rev.~Lett.}}    
\def\qjras{\aaref@jnl{QJRAS}}             
\def\skytel{\aaref@jnl{S\&T}}             
\def\ssr{\aaref@jnl{Space~Sci.~Rev.}}     
\def\zap{\aaref@jnl{ZAp}}                 
\def\nat{\aaref@jnl{Nature}}              
\def\aplett{\aaref@jnl{Astrophys.~Lett.}} 
\def\apspr{\aaref@jnl{Astrophys.~Space~Phys.~Res.}} 
\def\physrep{\aaref@jnl{Phys.~Rep.}}      
\def\physscr{\aaref@jnl{Phys.~Scr}}       
\def\commat{\aaref@jnl{Comm.~Math.~Phys.}}              
\def\science{\aaref@jnl{Science}}               
\def\cqg{\aaref@jnl{Classical Quant.~Grav.}}            
\def\jpcs{\aaref@jnl{JPCS}}                                     
\def\ijmpd{\aaref@jnl{Int.~J.~Mod.~Phys.~D}}                    
\def\grg{\aaref@jnl{Gen.~Relat.~Gravit.}}               
\def\rpp{\aaref@jnl{Rep.~Prog.~Phys.}}          
\def\npa{\aaref@jnl{Nucl.~Phys.~A}}        
\def\lrr{\aaref@jnl{Living Rev.~Rel.}}                   
\def\jcap{\aaref@jnl{J.~Cosmology Astropart.~Phys.}}    
\def\rmp{\aaref@jnl{Rev.~Mod.~Phys.}}   
\def\epjc{\aaref@jnl{Eur.~Phys.~J.~C}} 
\def\plb{\aaref@jnl{~Phy.~Lett.~B}} 
\def\mpla{\aaref@jnl{Mod.~Phy.~Lett.~A}} 
\def\arxiv{\aaref@jnl{arxiv.org}}

\allowdisplaybreaks[1]

\begin{document}

\color{black}   

\title{Constraining $f(T,\mathcal{T})$ Gravity with Gravitational Baryogenesis}

\author{Sai Swagat Mishra\orcidlink{0000-0003-0580-0798}}
\email{saiswagat009@gmail.com}
\affiliation{Department of Mathematics, Birla Institute of Technology and
Science-Pilani,\\ Hyderabad Campus, Hyderabad-500078, India.}

\author{Sanjay Mandal\orcidlink{0000-0003-2570-2335}}
\email{sanjaymandal960@gmail.com}
\affiliation{Department of Mathematics, Birla Institute of Technology and
Science-Pilani,\\ Hyderabad Campus, Hyderabad-500078, India.}

\author{P.K. Sahoo\orcidlink{0000-0003-2130-8832}}
\email{pksahoo@hyderabad.bits-pilani.ac.in}
\affiliation{Department of Mathematics, Birla Institute of Technology and
Science-Pilani,\\ Hyderabad Campus, Hyderabad-500078, India.}
%
\date{\today}
\begin{abstract}
Gravitational baryogenesis is one of the mechanisms which help us to explore more about our early universe, especially baryon-anti-baryon asymmetry. As we know, modified theories of gravity are very successful in describing the present accelerated scenario of the universe. Therefore, in this letter, we aim to constrain the generalized torsion-based modified theory of gravity, namely, $f(T,\mathcal{T})$ gravity with gravitational baryogenesis, where $T$, $\mathcal{T}$ are the torsion scalar, trace of the energy-momentum tensor, respectively. For this, we examine how the various Lagrangian forms of $f(T,\mathcal{T})$ affect the baryogenesis. We also impose the constraints on the extra degrees of freedom induced by modified theory with the observational values of the baryon-to-entropy ratio. In addition, we further explore how more generalized gravitational baryogenesis can attribute in a physically viable and consistent way to the cosmologies of the modified theory of gravity.\\

\end{abstract}

\maketitle

\date{\today}

\section{Introduction}

The physical process of Baryonic asymmetry that is hypothesized to happen in the early universe is known as Baryogenesis. The universe seems to have a positive baryon number density instead of zero, indicating an excess of matter over antimatter. Various predictions and observations like Big Bang Nucleosynthesis \cite{Burles/2001}, Cosmic Microwave Background \cite{{Bennett/2003},
{Spergel/2003}} have verified the predominance of matter over antimatter in the universe. To measure the asymmetry, baryon number density$(n_B)$ to entropy$(s)$ ratio$(n_B/s)$ can be used. BBN observed the ratio as $n_B/s=(5.6 \pm 0.6) \times 10^{-10}$ while CMB predicted as $n_B/s=(6.19 \pm 0.14) \times 10^{-10}$ at $95 \%$ CL. Numerically, the observed baryon-to-entropy ratio is $n_{B}/s=9.2^{+0.6}_{-0.4} \times 10^{-11}$. The imbalance of matter and antimatter is considered one of the most interesting mysteries among cosmologists. These articles \cite{{Rioto},{Dine},{Alexander/2006},{Mohanty/2006},{Li/2004},{Lambiase/2013},{Oikonomou/2016},{Oikonomou/2017},{Odintsov/2016},{Pizza}} can also be explored to get a detailed explanation of the asymmetry in different eras.

Sakharov proposed \cite{Sakharov/1967} a set of three conditions that are required to produce matter-antimatter asymmetry, namely 
(i) $B$ (Baryon number) Violation; 
(ii) $C$ (Charge) and $CP$ (Charge Parity) Violation; 
(iii) Interactions out of thermal equilibrium. Davoudiasl et al. \cite{Davoudiasl/2004} proposed a CP-violating interaction between the derivative of Ricci Scalar $R$ and the baryonic matter current $J^\mu$, in the form \\
\begin{equation}
\label{1}
    \frac{1}{M_*^2} \int \sqrt{-g} dx^4 (\partial_\mu R)J^\mu
\end{equation}
where $M_*$ represents the cutoff scale of the effective theory and $g$ is the metric determinant. This interaction dynamically breaks $CP$ during universe expansion and satisfies the second condition of Sakharov, which leads to Baryonic asymmetry. The baryon-to-entropy ratio $n_B/s$ is proportional to $\dot R$ if the interaction form \eqref{1} is applied in the case of flat Friedmann-Robertson-Walker (FRW) geometry. In the radiation-dominated phase $(\omega=1/3)$, the net baryon asymmetry induced by \eqref{1} is zero.

In this letter, we have discussed gravitational baryogenesis in $f(T, \mathcal{T})$ gravity which is a coupling between torsion scalar $(T)$ and trace of energy-momentum tensor $(\mathcal{T})$ respectively. Nowadays in modified gravitational theories, researchers prefer torsion-based teleparallel gravity instead of curvature-based General relativity. As the torsion formalism is equivalent to GR, it is also termed as the Teleparallel Equivalent of General Relativity (TEGR) \cite{{Unzicker/2005},{Moller/1961},{Hayashi/1979},{Arcos/2004},{Maluf/2013}}. The teleparallel gravity is generalized to $f(T)$ gravity, which can further be extended to $f(T, \mathcal{T})$ gravity \cite{{Harko/2014},
{Mandal/2023},
{Arora/2022}}. The $f(T, \mathcal{T})$ gravity is pretty different from other Torsion or curvature-based models. It has some exciting features like describing the history of expansion with an initial inflationary phase, a following matter-dominated expansion without acceleration, and at last late-time accelerating phase \cite{Momeni/2014}. It has also been investigated in various areas such as quark stars \cite{Pace/2017}, reconstruction and stability \cite{Junior/2016}, and the growth factor of sub-horizon modes \cite{Farrugia/2016}.

In particular, we shall study the properties of the gravitational baryogenesis terms in detail corresponding to $\partial_{\mu}(T+\mathcal{T})$ or $\partial_{\mu}\,f(T,\mathcal{T})$. We shall discuss the resulting baryon-to-entropy ratios for each model that can be aligned with its' observational values. This is only possible if we choose some model parameter's value arbitrarily large or too small. Furthermore, we shall constrain the functional forms of Lagrangian $f(T,\mathcal{T})$ for generalized baryogenesis cases and discuss their cosmologies.

This letter is organized as follows: We initiate by reviewing the basic equations of the torsion-based gravity and solving the motion equations for the solution of the Hubble parameter in section \ref{sec2}. Then we investigate several gravitational baryogenesis scenarios in terms of $f(T,\mathcal{T})$ gravity by assuming various linear and nonlinear models in section \ref{sec3} and \ref{sec4}. In the end, we conclude the results.

\section{Basic Formalism of $f(T,\mathcal{T})$ gravity} \label{sec2}
Teleparallel gravity uses the curvature-free Weitzenb $\ddot{o}$ck connection, instead of the Torsion-free Levi-Civit\'{a} connection of GR. The weitzenb$\ddot{o}$ck connection can be defined as $\overset{w}{\Gamma}_{\nu \mu}^\lambda \equiv e_{A}^\lambda\partial_{\mu}e_{\nu}^A$ and the metric tensor can be expressed as $g_{\mu \nu}=\eta_{AB}e_{\mu}^A e_{\nu}^B$. Here $e_{A}^\lambda$ and $e_{\mu}^A$ are the tetrads (vierbeins) and $\eta_{AB}$ is the Minkowski metric tensor $(\eta_{AB}=diag(1,-1,-1,-1))$.\\

Using the above connection, the Torsion tensor can be defined as
\begin{equation}
\label{2}
    T_{\mu \nu}^\lambda=\overset{w}{\Gamma}_{\nu \mu}^\lambda-\overset{w}{\Gamma}_{\mu \nu}^\lambda.
\end{equation}
 The superpotential tensor can be expressed as,
\begin{equation}
\label{3}
    {S_{\rho}}^{\mu \nu} \equiv \frac{1}{2}({K^{\mu \nu}}_{\rho}+\delta_\rho^\mu {T^{\alpha \nu}}_{\alpha}-\delta_\rho^\nu {T^{\alpha \mu}}_{\alpha}).
\end{equation}
where ${K^{\mu \nu}}_{\rho}$ is the contorsion tensor which can be expressed in terms of Torsion tensor as ${K^{\mu \nu}}_{\rho} \equiv -\frac{1}{2}({T^{\mu \nu}}_{\rho}-{T^{\nu \mu}}_{\rho}-{T_{\rho}}^{\mu \nu})$.
Combining both the equations \eqref{2} and \eqref{3} we can get the Torsion scalar T,
\begin{equation}
\label{4}
    T\equiv  {S_{\rho}}^{\mu \nu} T_{\mu \nu}^\rho
      = \frac{1}{4}{T^{\rho \mu \nu}T_{\rho \mu \nu}}+ \frac{1}{2}{T^{\rho \mu \nu}T_{\nu \mu \rho}}-{T_{\rho \mu}}^\rho {T^{\nu \mu}}_\nu,
\end{equation}
which defines the action of teleparallel gravity as,
\begin{equation}
\label{5}
    S=\frac{1}{16\pi G}{\int{{d^4}xeT+\int{{d^4}xe\mathcal{L}_m}}},
\end{equation}
where $e=det(e_\mu^A)=\sqrt{-g}$. $G$ and $\mathcal{L}_m$ represents 
 the Newton's constant and matter Lagrangian respectively. Further, the torsion scalar in the above equation can be generalized to a function of both the torsion scalar and the trace of energy-momentum tensor as $T+f(T,\mathcal{T})$. So the revised action for $f(T,\mathcal{T})$ gravity can be expressed as,
 \begin{equation}
\label{6}
  S=\frac{1}{16\pi G}{\int{{d^4}xe[T+f(T,\mathcal{T})]+\int{{d^4}xe\mathcal{L}_{m}}}},
\end{equation}
Now, we consider a spatially flat Friedmann-Lemaitre-Robertson-Walker (FLRW) metric to analyze the geometric description of the universe,
   \begin{equation}
   \label{7}
       ds^2=dt^2-a^2(t)\delta_{ij}dx^i dx^j,
   \end{equation}
   where a(t) is the scale factor as a function of time. For the above metric, the vierbein field can be expressed as $e_\mu^A=diag(1,a,a,a)$.\\
By varying the action\eqref{6} with respect to the inverse vierbein field, we get the following field equations
\begin{multline}
 \label{8}
   (1+f_T)\left[e^{-1}\partial_\sigma(e\,{e_A}^\alpha S_\alpha^{\rho\sigma})-e_A^\alpha T^{\sigma}_{\nu \alpha} S_{\sigma}^{\nu \rho} \right]+
   \large(f_{TT}\, \partial_{\sigma} T+ \\
   f_{T \mathcal{T}}\,\partial_{\sigma} \mathcal{T}\large)e\,e_{A}^{\alpha} S_{\alpha}^{\rho \sigma}+e_{A}^\rho\left(\frac{f+T}{4}\right)\\
-\frac{f_{\mathcal{T}}}{2} \left(e^{\alpha}_A \stackrel{em}{T}_\alpha^{\rho} +p\, e_A^{\rho} \right)= 4\pi G\, e^{\alpha}_A \stackrel{em}{T}_\alpha^{\rho},
\end{multline}
 where ${\stackrel{em}{T}_\alpha}^\rho$ is the stress-energy tensor, $T=-6 H^2$, $\mathcal{T}= \rho_m-3p_m$ which holds in perfect matter fluid case,$f_T={\partial f}/{\partial T}$ and $f_{T\mathcal{T}}={\partial^2{f}}/{\partial T \partial \mathcal{T}}$.

Using the above FLRW metric\eqref{7} in the field eq. \eqref{8}, we obtain the modified Friedmann equations:
\begin{equation}\
\label{9}
  {H^2=\frac{8\pi G}{3}{\rho_m}-\frac{1}{6}(f+12H^2f_T)+f_\mathcal{T}(\frac{\rho_m+p_m}{3})},
 \end{equation}
  \begin{multline}
  \label{10}
  {\dot{H}=-4\pi G(\rho_m+p_m)}-\dot{H}(f_T-12H^2f_{TT})\\-H(\dot{\rho_m}-3{\dot{p_m}})f_{T\mathcal{T}}-f_\mathcal{T}(\frac{\rho_m+p_m}{2}).
  \end{multline}
 We will use the above field equations to obtain the Baryon-to-entropy ratio for different models. 
  
\section{Baryogenesis in $f(T,\mathcal{T})$ Gravity} \label{sec3}

For $f(T,\mathcal{T})$ gravity, we consider a $CP$-Violating interaction term generated by the baryonic asymmetry of the form,
\begin{equation}
\label{11}
    \frac{1}{M_*^2} \int \sqrt{-g} dx^4 (\partial_\mu(-T-\mathcal{T}))J^\mu.
\end{equation}
In our further work, we also assume that thermal equilibrium exists with energy density being proportional to the decoupling temperature $T_D^*$ as
\begin{equation}
\label{12}
    \rho = \frac{\pi^2}{30} g_* (T_D^*)^4,
\end{equation}
where $g_*= \frac{45s}{2 \pi^2 (T_D^*)^3}$ represents the number of degrees of freedom of the particles contributing to the universe's global entropy. For the above $CP$-Violating interaction term\eqref{11}, the induced chemical potential can be written as $\mu\sim \pm \frac{(\dot{T}+\dot{\mathcal{T}})}{M_*^2}$, hence the corresponding Baryon-to-entropy ratio defined as
\begin{equation}
\label{13}
    \frac{n_B}{S} \simeq -\frac{15g_b}{4\pi^2 g_*}\left[\frac{1}{M_*^2 T^*}(\dot{T} + \dot{\mathcal{T}})\right]_{T^*=T_D^*}, 
\end{equation}
where $g_b$ is the total number of intrinsic degrees of freedom of baryons.

For the calculation of the baryon-to-entropy ratio, we have considered the power-law cosmic evolution with scale factor as
\begin{equation}
\label{14}
    a(t)=At^n,
\end{equation}
where $A$ and $n$ are positive constants. The consistency of the assumed scale factor can be verified in the first motion equation \eqref{9} along with the respective form of Lagrangian $f(T,\mathcal{T})$ in radiation dominated phase. The values of free parameters $A$ and $n$ may vary from one model to another to obtain desired results, but in our study, we have investigated precisely with the exact solution to the scale factor for respective $f(T,\mathcal{T})$ form in the coming sections. In GR the baryon-to-entropy ratio is equal to zero during the radiation-dominated phase\cite{Bennett/2003}. However, in some modified theories of gravity, it may be nonzero\cite{{Oikonomou},{Baffou/2019},
{Sahoo/2020},{Bhattacharjee/2020}}. To verify this in  $f(T, \mathcal{T})$ we have assumed $\omega=1/4$ in all the models and their generalized form. In the case of $\omega=1/3$, the gravity simply reduces to $f(T)$ which has already been explored in gravitational baryogenesis\cite{Oikonomou}.

\subsection{Model-I}

Let's consider a linear form of Lagrangian as
\begin{equation}
\label{15}
    f(T, \mathcal{T})=\alpha T + \beta \mathcal{T},
\end{equation}
$\alpha$ and $\beta$ being free parameters. \\

The analytic form of the scale factor \eqref{14} can be obtained from \eqref{9} and \eqref{15} as $a(t)=\sqrt{2\gamma}t^{1/2}$, where $\gamma=\frac{\rho_0 (3+4\beta)}{9(1+\alpha)}$ which is a constant.\\

Now, using the above scale factor and lagrangian form of $f(T,\mathcal{T})$ in the first motion equation \eqref{9} we get the analytic expression of energy density as,

\begin{equation}
\label{16}
    \rho=\frac{\delta}{ t^2},
\end{equation}

where $\delta=\frac{3(\alpha +1)}{2(2+\beta(1+5\omega))}$.
Equating the above equation\eqref{16} with \eqref{12}, we get the decoupling time $t_D$ in terms of the decoupling temperature $T_D^*$ as
\begin{equation}
\label{17}
   t_D= \frac{1}{\pi {T_D^*}^2 }
   \sqrt{\frac{30 \delta}{g_*}}.
\end{equation}
Using \eqref{13} we obtain the baryon-to-entropy ratio for this model as,
\begin{multline}
    \label{18}
    \frac{n_B}{S} \simeq -\frac{15 g_b}{4 \pi^2 g_* M_*^2 T_D^*}
    \left(\frac{3-2 \delta +6 \delta \omega}{\left(\frac{1}{\pi {T_D^*}^2 }
   \sqrt{\frac{30 \delta}{g_*}}\right)^3}\right).
\end{multline}

 \begin{figure}[H]
    \centering
    \includegraphics[scale=0.77]{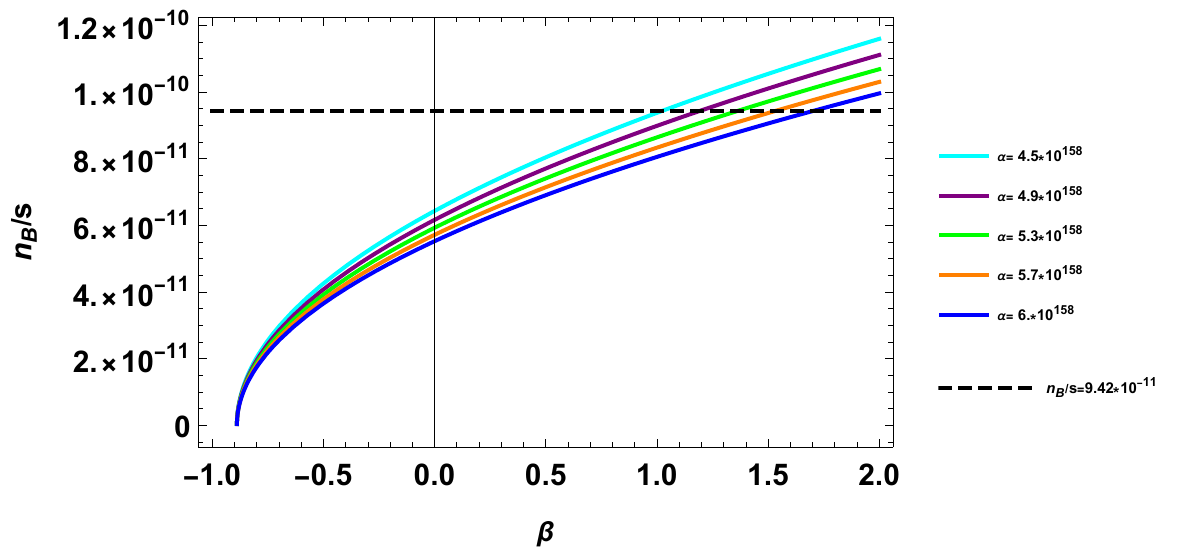}
    \caption{The $\beta$-dependence of baryon-to-entropy ratio for Model-I for varying $\alpha$ with $g_b=1$, $T_D^*=2\times10^{16} GeV$, $M_*=10^{12} GeV$, $g_*=106$ and $\omega=\frac{1}{4}$.}
     \label{fig 1}
\end{figure}

In fig.\ref{fig 1}, we plot the curves by fixing five different values of $\alpha$ which are represented by different colors, while the rest values are the same for each curve. There should not be any confusion between our model parameters and the assumed parameter $\delta$. As $\delta$ is an expression with both $\alpha$ and $\beta$ and, one can obtain $\delta$ by substituting the values. The dotted line is the intersection of all the curves at $n_B/s=9.42\times 10^{-11}$. We observe that for certain fixed values to our free parameters, this model agrees with the observed value of the baryon-to-entropy ratio.

\subsection{Model-II}

Let's consider a non-linear form of Lagrangian with the square of $\mathcal{T}$ as
\begin{equation}
\label{19}
    f(T, \mathcal{T})=\alpha T + \beta \mathcal{T}^2.
\end{equation}

The analytic form of scale factor can be obtained from \eqref{9} and \eqref{19} as $a(t)=\sqrt{2\zeta}t^{1/2}$, where $\zeta=\frac{\rho_0}{3(1+\alpha)}$ which is a constant.\\
Now, using the above scale factor and lagrangian form of $f(T,\mathcal{T})$ in the first motion equation \eqref{9} we get the analytic expression of energy density as,

\begin{equation}
\label{20}
\rho=\frac{-\frac{1}{3} + \sqrt{\frac{1}{9} +\frac{\delta_1(1+\alpha)}{t^2}}}{2\delta_1},
\end{equation}
where $\delta_1 =\frac{\beta}{6} (1-3 \omega)(3+7 \omega)$.
Equating the above equation with the energy density \eqref{12}, we get the decoupling time $t_D$ in terms of the decoupling temperature $T_D^*$ as
\begin{equation}
\label{21}
    t_D= \frac{1}{2 \pi {T_D^*}^2} \sqrt{\frac{1+ \alpha}{\frac{g_*}{30} \left(\frac{1}{3} + \frac{\pi^2}{30} g_* {T_D^*}^4 \delta_1 \right)}}.
\end{equation}
Using \eqref{13} we obtain the baryon-to-entropy ratio for this model as,
\begin{multline}
    \label{22}
    \frac{n_B}{S} \simeq -\frac{ g_b \pi {T_D^*}^5 }{2 \sqrt{30}  g_* M_*^2 } \\ \times
    \left(\frac{6+ \frac{15 (1+ \alpha ) (-1+ 3 \omega )}{5+\pi ^2 \delta_1  g_* {T_D^*}^4 }}{  \left(\frac{1+ \alpha}{g_* \left(\frac{\pi ^2}{30}  \delta_1  g_* {T_D^*}^4+ \frac{1}{3} \right)}\right)^{3/2}} \right).
 \end{multline} 
 
 \begin{figure}[H]
    \centering
    \includegraphics[scale=0.77]{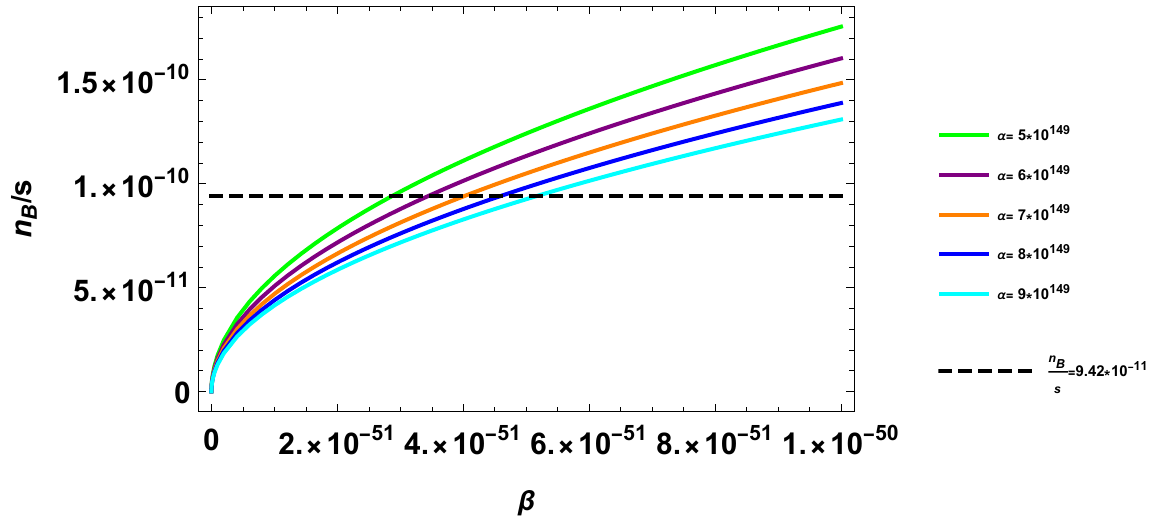}
    \caption{The $\beta$-dependence of baryon-to-entropy ratio for Model-II for varying $\alpha$ with $g_b=1$, $T_D^*=2\times10^{16} GeV$, $M_*=10^{12} GeV$, $g_*=106$ and $\omega=\frac{1}{4}$}
    \label{fig2}
\end{figure}
In fig.\ref{fig2} we plot the curves with five different values of $\alpha$, while the rest values are same for each curve. The intersection at $n_B/s=9.42 \times 10^{-11}$ is represented by the dashed line. Hence Model-II agrees with the observed numerical value.

 \subsection{Model-III}
 
 Let's consider a non-linear form of Lagrangian with $r$ power of the Torsion scalar as
\begin{equation}
\label{23}
    f(T, \mathcal{T})=\alpha T^r + \beta \mathcal{T},
\end{equation}
with $r$ being a positive constant. Using the assumed scale factor\eqref{14} and the above model \eqref{23} in the first motion equation \eqref{9} we obtain the energy density at leading order is,
\begin{equation}
\label{24}
    \rho \simeq Ct^{-2r},
\end{equation}
where $C= \frac{\alpha (1-2r) (-6n^2)^r}{2+ \beta + 5\beta \omega }$.
Equating the above expression with the energy density \eqref{12}, we get the decoupling time $t_D$ in terms of the decoupling temperature $T_D^*$ as
\begin{equation}
\label{25}
 t_D= \left(\frac{\pi^2 g_*}{30C}\right)^{-{\frac{1}{2r}}} {T_D^*}^{-\frac{2}{r}}.
\end{equation}
Using \eqref{13} we obtain the baryon-to-entropy ratio for this model as,
\begin{multline}
    \label{26}
    \frac{n_B}{S} \simeq -\frac{15 g_b}{4 \pi^2 g_* M_*^2 T_D^*}\\ \times
    \left(12n^2 \left(\frac{\pi^2 g_*}{30C}\right)^{\frac{3}{2r}} {T_D^*}^{\frac{6}{r}} - 2Cr(1-3 \omega) \left(\frac{\pi^2 g_*}{30C}\right)^{1+ \frac{1}{2r}} {T_D^*}^{4+ \frac{2}{r}}\right)
  \end{multline}  
  
  \begin{figure}[H]
    \centering
    \includegraphics[scale=0.77]{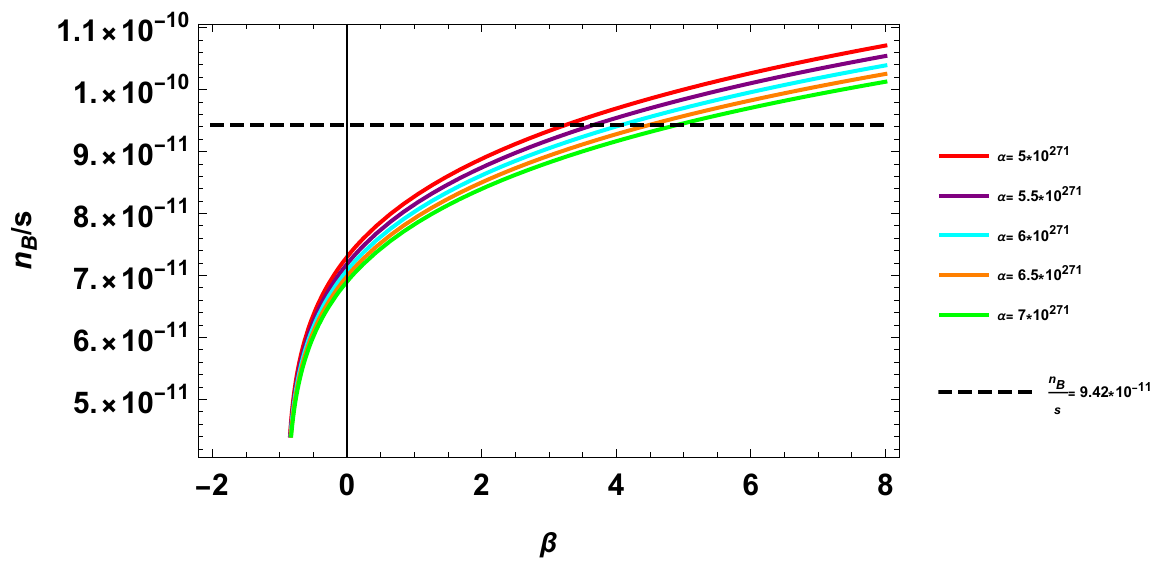}
    \caption{The $\beta$-dependence of baryon-to-entropy ratio for Model-III for varying $\alpha$ with $g_b=1$, $T_D^*=2\times10^{16} GeV$, $M_*=10^{12} GeV$, $g_*=106$,  $r=3$, $n=\frac{1}{2}$ and $\omega=\frac{1}{4}$.}
    \label{fig3}
\end{figure}

Similar to Model-I and II, in fig.\ref{fig3} also we have obtained the baryon-to-entropy ratio that agrees with the observed value, by varying five values of $\alpha$ and fixing the rest values. Though we have fixed the value of $r=3$ here, it can be increased to any positive number to obtain the desired value by changing the free parameters accordingly.

\begin{widetext}

\begin{table}[H]
 \centering
 \caption{
 Change in the model parameters $\alpha$ and $\beta$ to obtain the observed baryon-to-entropy ratio.}
 
 \label{table1}
    \begin{tabular}{||c||c|c|c||}
    \hline
    Model & $\alpha$ & $\beta$ & $n_B/s$ \\
    \hline
      & $10^{159}$ & $3.36$ & $9.42878 \times 10^{-11}$ \\
    
   $f(T,\mathcal{T})=\alpha T + \beta \mathcal{T}$  & $10^{160}$ & $41.56$ & $9.42434 \times 10^{-11}$ \\
    
    & $10^{161}$ & $423.36$ & $9.42168 \times 10^{-11}$ \\
    \hline
    \hline
    Model & $\alpha$ & $\beta$ & $n_B/s$ \\
    \hline
      & $10^{141}$ & $57.61 \times 10^{-61}$ & $9.42029 \times 10^{-11}$ \\
    
   $f(T,\mathcal{T})=\alpha T + \beta \mathcal{T}^2$ &  $10^{142}$ & $57.65 \times 10^{-60}$ & $9.42356 \times 10^{-11}$ \\
    
    & $10^{143}$ & $57.7 \times 10^{-59}$ & $9.42765 \times 10^{-11}$ \\
    
    \hline
    \hline
    Model & $\alpha$ & $\beta$ & $n_B/s$  \\ \hline
       & $10^{273}$ & $82$ & $9.42825 \times 10^{-11}$ \\
    
   $f(T,\mathcal{T})=\alpha T^r + \beta \mathcal{T}$ &  $10^{274}$ & $825$ & $9.42256 \times 10^{-11}$ \\
    
    & $10^{275}$ & $8250$ & $9.42103 \times 10^{-11}$ \\
    \hline
    \end{tabular}
\end{table}

\end{widetext}

In Table \ref{table1}, our motivation is to keep matching the value of the baryon-to-entropy ratio with the numerically observed value and analyze the change in model parameters $\alpha$ and $\beta$. In all three models, it is evident that $\alpha$ and $\beta$ increase together. One can observe from the table and plot that the list of values of model parameters is not exhaustive. We can constrain the values with different ranges to obtain the numerical result $n_B/s \sim 9.42 \times 10^{-11}$. For calculation, we have fixed the values $g_b=1$, $T_D^*=2\times10^{16} GeV$, $M_*=10^{12} GeV$, $g_*=106$,  $r=3$ and $\omega=\frac{1}{4}$.

\section{Generalized Baryogenesis Term}\label{sec4}

 Here we consider a more general baryogenesis interaction term by extending \eqref{11} to,
 \begin{equation}
     \label{27}
    \frac{1}{M_*^2} \int \sqrt{-g} dx^4 (\partial_\mu(-T+ -\mathcal{T} +f(T, \mathcal{T}))\,J^\mu.
 \end{equation}
We have already calculated the $-T- \mathcal{T}$ part in the previous subsection. So we are interested in calculating the remaining part here. To obtain the full $-T- \mathcal{T} + f(T, \mathcal{T})$ result, one can add both the values of the baryon-to-entropy ratio by calculating them individually.
We define the Baryon to entropy ratio as,
\begin{equation}
\label{28}
    \frac{n_B}{S} \simeq -\frac{15g_b}{4\pi^2 g_*}\left[\frac{1}{M_*^2 T^*}(\dot{T} f_T (T, \mathcal{T}) +\dot{\mathcal{T}} f_\mathcal{T}(T, \mathcal{T}))\right]_{T^*=T_D^*} .
\end{equation}

\begin{itemize}
    \label{29}
    {\bfseries
    \item For Model-I the resulting baron-to-entropy ratio is,
    \begin{multline}
       \frac{n_B}{S} \simeq -\frac{15 g_b}{4 \pi ^2 g_* M_*^2 T_D^*}\left( \frac{3 \alpha +6 \beta  \delta  \omega -2 \beta  \delta }{\left(\frac{1}{\pi  T_D^{*2}} \sqrt{\frac{30 \delta }{g_*}} \right)^3} \right),
    \end{multline}
    
    where $\delta=\frac{3(\alpha +1)}{2(2+\beta(1+5\omega))}$.
    \item For Model-II the resulting baron-to-entropy ratio is,
    \begin{multline}
         \label{30}
       \frac{n_B}{S} \simeq -\frac{15 g_b \pi {T_D^*}^5}{ g_* M_*^2}  \times
       \left( 6- \frac{30 \beta \rho (1+ \alpha)(1-3 \omega)^2}{5+g_* {T_D^*}^4 \pi^2 \delta_1 }\right) \\  \times \left(\frac{g_* \left(\frac{\pi^2}{30} g_* {T_D^*}^4  \delta_1 + \frac{1}{3} \right)}{30(1+ \alpha)} \right)^\frac{3}{2},
    \end{multline}
    where $\rho=\frac{-\frac{1}{3} + \sqrt{\frac{1}{9} +\frac{\delta_1(1+\alpha)}{t^2}}}{2\delta_1}$ and $\delta_1 =\frac{\beta}{6} (1-3 \omega)(3+7 \omega)$.
    }
    \item For Model-III the resulting baron-to-entropy ratio is,
   \begin{multline}
         \label{31}
        \frac{n_B}{S} \simeq -\frac{15 g_b}{4 \pi^2 g_* M_*^2 T_D^*} 2r {T_D^*}^{4+ \frac{2}{r}} \\  \times \left(C \beta (1-3 \omega)+ (-6n^2)^r \alpha \right) \left(\frac{\pi^2 g_*}{30C}\right)^{1+ \frac{1}{2r}} ,
   \end{multline}
   where $C= \frac{\alpha (1-2r) (-6n^2)^r}{2+ \beta + 5\beta \omega }$.
\end{itemize}

\begin{figure}[H]
    \centering
    \includegraphics[scale=0.77]{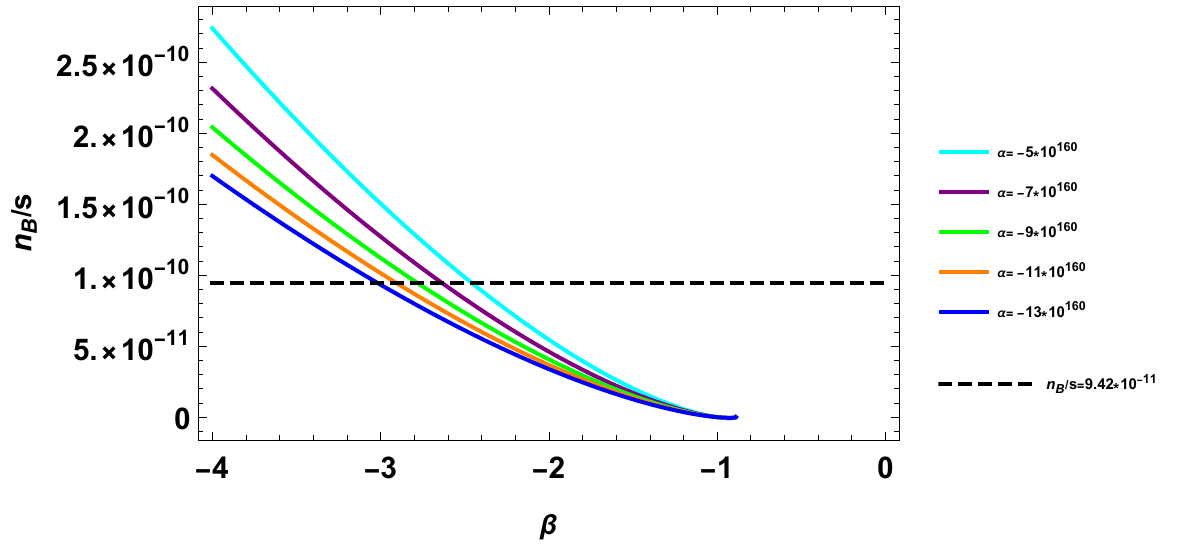}
    \caption{The $\beta$-dependence of baryon-to-entropy ratio for the generalized case of Model-I for varying $\alpha$ with $g_b=1$, $T_D^*=2\times10^{16} GeV$, $M_*=10^{12} GeV$, $g_*=106$ and  $\omega=\frac{1}{4}$.}
    \label{fig4}
\end{figure}
\begin{figure}[H]
    \centering
    \includegraphics[scale=0.77]{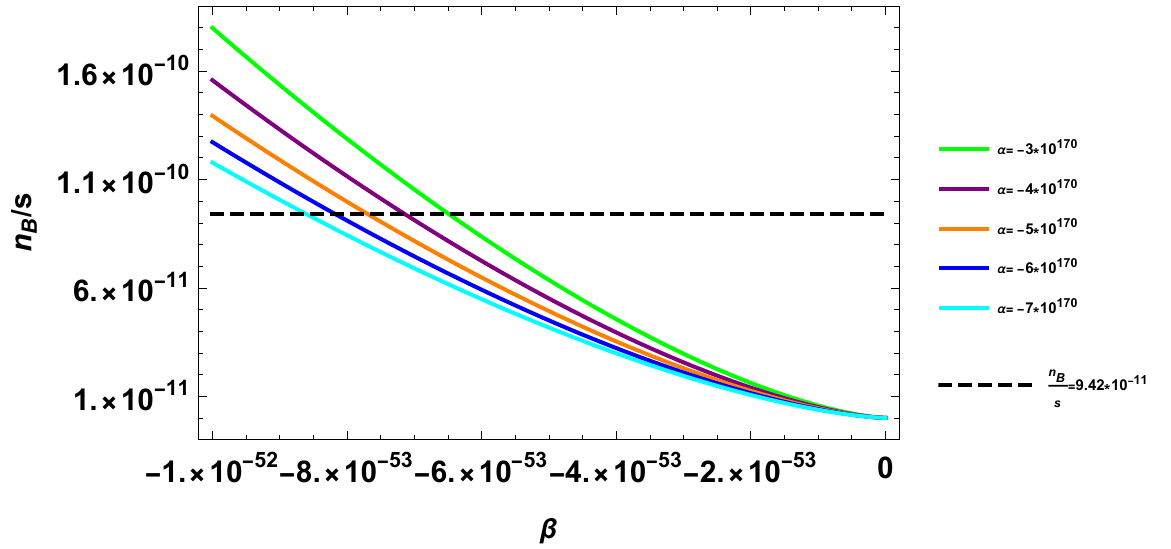}
    \caption{The $\beta$-dependence of baryon-to-entropy ratio for the generalized case of Model-II for varying $\alpha$ with $g_b=1$, $T_D^*=2\times10^{16} GeV$, $M_*=10^{12} GeV$, $g_*=106$ and $\omega=\frac{1}{4}$.}
    \label{fig5}
\end{figure}
\begin{figure}[H]
    \centering
    \includegraphics[scale=0.77]{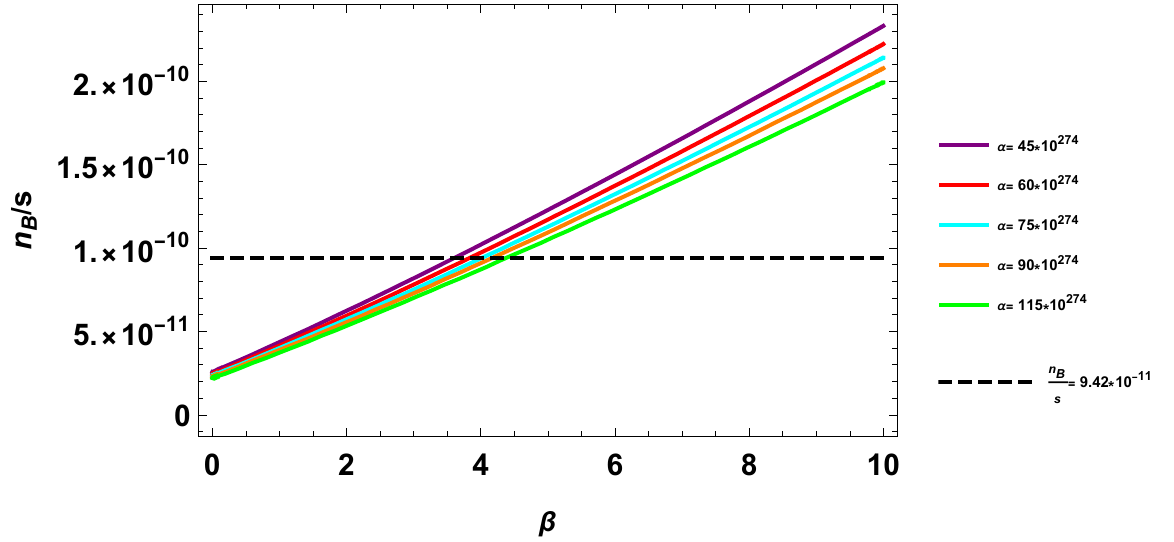}
    \caption{The $\beta$-dependence of baryon-to-entropy ratio for the generalized case of Model-III for varying $\alpha$ with $g_b=1$, $T_D^*=2\times10^{16} GeV$, $M_*=10^{12} GeV$, $g_*=106$, $r=3$, $n=\frac{1}{2}$ and $\omega=\frac{1}{4}$.}
    \label{fig6}
\end{figure}

The values of $\delta$ and $\delta_1$ can be obtained by substituting values of the model parameters in fig.\ref{fig4} and \ref{fig5} respectively, $8 \pi G$ is considered $1$. Similarly, $C$ can be obtained by substituting the respective values in fig.\ref{fig6}. Hence the generalized cases of all the assumed models agree with the observed value for fixed values of the free parameters.

\begin{widetext}

\begin{table}[H]
 \centering
 \caption{
 Change in the model parameters $\alpha$ and $\beta$ to obtain the observed baryon-to-entropy ratio.}
 
 \label{table2}
    \begin{tabular}{||c||c|c|c||}
    \hline
    Model & $\alpha$ & $\beta$ & $n_B/s$ \\
    \hline
      & $-1.112 \times 10^{163}$ & $-10$ & $9.42722 \times 10^{-11}$ \\
    
    Generalized term for $f(T,\mathcal{T})=\alpha T + \beta \mathcal{T}$  & $-0.1525 \times 10^{164}$ & $-11$ & $9.42263 \times 10^{-11}$ \\
    
    & $-0.02028 \times 10^{165}$ & $-12$ & $9.42205 \times 10^{-11}$ \\
    \hline
    \hline
    Model & $\alpha$ & $\beta$ & $n_B/s$ \\
    \hline
      & $-0.10935 \times 10^{184}$ & $-10^{-48}$ & $9.42373 \times 10^{-11}$ \\
    
   Generalized term for $f(T,\mathcal{T})=\alpha T + \beta \mathcal{T}^2$ &  $-0.1094 \times 10^{181}$ & $-10^{-49}$ & $9.42158 \times 10^{-11}$ \\
    
    & $-0.1093 \times 10^{178}$ & $-10^{-50}$ & $9.42589 \times 10^{-11}$ \\
    
    \hline
    \hline
    Model & $\alpha$ & $\beta$ & $n_B/s$  \\ \hline
       & $84 \times 10^{279}$ & $29$ & $9.42624 \times 10^{-11}$ \\
    
   Generalized term for $f(T,\mathcal{T})=\alpha T^r + \beta \mathcal{T}$ &  $105 \times 10^{279}$ & $30$ & $9.42665 \times 10^{-11}$ \\
    
    & $1305 \times 10^{278}$ & $31$ & $9.4252 \times 10^{-11}$ \\
    \hline
    \end{tabular}
\end{table}

\end{widetext}
In Table \ref{table2}, we have performed the same analysis as Table \ref{table1}. For calculation, we have fixed the values $g_b=1$, $T_D^*=2\times10^{16} GeV$, $M_*=10^{12} GeV$, $g_*=106$,  $r=3$ ,$n=\frac{1}{2}$ and $\omega=\frac{1}{4}$.

\section{Conclusion}\label{sec5}

In this letter, we have investigated the gravitational baryogenesis thoroughly in the framework of $f(T, \mathcal{T})$ gravity. We find that the baryon-to-entropy ratio for the $CP$-violating interaction term is proportional to $\partial_ \mu (-T- \mathcal{T})$ and $\partial_\mu f(T, \mathcal{T})$. By choosing the lagrangian form $ f(T, \mathcal{T})=\alpha T + \beta \mathcal{T}$, $ f(T, \mathcal{T})=\alpha T + \beta \mathcal{T} ^2$ and  $f(T, \mathcal{T})=\alpha T^r + \beta \mathcal{T}$, we observe that one can easily achieve the baryon-to-entropy ratio that agrees with the observed value $n_B/s= 9.42 \times 10^{-11}$. For that, we have just assigned some suitable values to our free parameters. In all the calculations we have assumed $g_b=1$, $T_D^*=2\times10^{16} GeV$, $M_*=10^{12} GeV$ and $g_*=106$. The equation of state parameter $(\omega)$ is assumed as $1/4$, which is very close to the radiation  era. Finally, we have observed the baryon asymmetry in more generalized cases and found that the same process leads us to the desired result. Furthermore, we are able to put constraints on the free parameters of our models through the observational constraint values of the baryon-to-entropy ratio. We also observed that our model's baryon-to-entropy ratios are aligned with its' observational values either only for very large values or very tiny values of free parameters. In addition, we showed that the studied models are consistent with the observed data and do not vanish $n_B/s$, whereas, in the TEGR(Teleparallel Equivalent to General Relativity), formulation inconsistency occurs \cite{ft}. With this method, one can explore more general cosmological evolution, and the resulting baryon-to-entropy ratio is compatible with the observational data.

\section*{Data Availability Statement}
There are no new data associated with this article.

\section*{Acknowledgements}
 SSM acknowledges the Council of Scientific and Industrial Research (CSIR), Govt. of India  for awarding Junior Research fellowship (E-Certificate No.: JUN21C05815). PKS  acknowledges the Science and Engineering Research Board, Department of Science and Technology, Government of India for financial support to carry out the Research project No.: CRG/2022/001847. We are very much grateful to the honorable referees and to the editor for the illuminating suggestions that have significantly improved our work in terms of research quality, and presentation.


\begin{thebibliography}{90}

\bibitem{Burles/2001} S. Burles, K.M. Nollett, M.S. Turner, Phys. Rev. D \textbf{63}, 063512 (2001). arXiv:astro-ph/0008495

\bibitem{Bennett/2003} C. Bennett, Ap. J. S. \textbf{148}, 1 (2003)

\bibitem{Spergel/2003} D. Spergel et al., Ap. J. S. \textbf{148}, 175 (2003)

\bibitem{Rioto} A. Rioto, M. Trodden's, arXiv:hep-ph/9901362

\bibitem{Dine} M. Dine, A. Kusenko, arXiv:hep-ph/0303065

\bibitem{Alexander/2006} S.H.S. Alexander, M.E. Peskin, M.M. Sheikh-Jabbari, Phys. Rev. Lett. \textbf{96}, 081301 (2006)

\bibitem{Mohanty/2006} S. Mohanty, A.R. Prasanna, G. Lambiase, Phys. Rev. Lett. \textbf{96}, 071302 (2006)

\bibitem{Li/2004} H. Li, M. Li, X. Zhang, Phys. Rev. D \textbf{70}, 047302 (2004).
arXiv:hep-ph/0403281

\bibitem{Lambiase/2013} G. Lambiase, S. Mohanty, A.R. Prasanna, Int. J. Mod. Phys. D \textbf{22}, 1330030 (2013). arXiv:1310.8459v1 

\bibitem{Oikonomou/2016} V.K. Oikonomou, Int. J. Geom. Methods Mod. Phys. \textbf{13},
1650033 (2016). arXiv:1512.04095v2 

\bibitem{Oikonomou/2017} V.K. Oikonomou, Supriya Pan, Rafael C. Nunes, Int. J. Mod. Phys.
A \textbf{32}, 1750129 (2017). arXiv:1610.01453v1

\bibitem{Odintsov/2016} S. D. Odintsov, V. K. Oikonomou, EPL \textbf{116}, 49001 (2016).
arXiv:1610.02533

\bibitem{Pizza} L. Pizza, arXiv:1506.08321

\bibitem{Sakharov/1967} A.D. Sakharov, JETP Lett. \textbf{5}, 24 (1967)

\bibitem{Davoudiasl/2004} H. Davoudiasl, R. Kitano, G.D. Kribis, H. Murayama, P. Steinhardt, Phys. Rev. Lett. \textbf{93}, 201301 (2004)

\bibitem{Unzicker/2005} A. Unzicker, T. Case, arXiv:physics/0503046.

\bibitem{Moller/1961} C. Moller, Conservation laws and absolute parallelism in general relativity, Mat-Fys. Skr. Udg. K. Da. \textbf{1},  3 (1961).

\bibitem{Hayashi/1979} K. Hayashi, T. Shirafuji, Phys. Rev. D \textbf{19}, 3524  (1979).

\bibitem{Arcos/2004} H.I. Arcos, J.G. Pereira, Int. J. Mod. Phys. D \textbf{13}, 2193 (2004).

\bibitem{Maluf/2013} J.W. Maluf,  Annalen Phys. \textbf{525}, 339 (2013).

\bibitem{Harko/2014} Harko, et al., J. Cosmol. Astropart. Phys. \textbf{12}, 021 (2014)

\bibitem{Mandal/2023} S. Mandal, S. S. Mishra, P. K. Sahoo.	arXiv:2301.06328. (2023)

\bibitem{Arora/2022} S. Arora, A. M. D.  Bhat, P.K. Sahoo. arXiv:2210.01552.

\bibitem{Momeni/2014}  D. Momeni, R. Myrzakulov, Int. J. Geom. Methods Mod. Phys. \textbf{11}, 1450077 (2014)

\bibitem{Pace/2017} M. Pace, J. Levi Said, Eur. Phys. J. C \textbf{77}, 62 (2017).

\bibitem{Junior/2016}  E. L. B. Junior, M. E. Rodrigues, I. G. Salako, M. J. S. Houndjo, Class.l Quantum Grav. \textbf{33}, 125006 (2016).

\bibitem{Farrugia/2016}  G. Farrugia, J. Levi Said, Phys. Rev. D \textbf{94}, 124004 (2016).

\bibitem{Oikonomou} V.K. Oikonomou, Emmanuel N. Saridakis, Phys. Rev. D \textbf{94},
124005 (2016). arXiv:1607.08561v2

\bibitem{Baffou/2019} Baffou, E.H., Houndjo, M.J.S., Kanfon, D.A. et al. Eur. Phys. J. C \textbf{79}, 112 (2019).

\bibitem{Sahoo/2020} Sahoo, Bhattacharjee, Int J Theor Phys \textbf{59}, 1451 (2020).

\bibitem{Bhattacharjee/2020} Bhattacharjee, Sahoo,  Eur. Phys. J. C \textbf{80}, 289 (2020).

\bibitem{ft} V.K. Oikonomou, Emmanuel N. Saridakis, Phys. Rev. D. \textbf{94}, 124005 (2016).

\end{thebibliography}
\end{document}